\documentclass[useAMS,usenatbib,usegraphicx]{mn2e}

\usepackage{ amssymb }
\usepackage{ hhline }

\title[Tides, planetary companions, and habitability]{Tides, planetary companions, and habitability: Habitability in the habitable zone of low-mass stars}
\author[C. Van Laerhoven, R. Barnes, and R. Greenberg]{C. Van Laerhoven$^{1}$\thanks{E-mail:
cvl@lpl.arizona.edu or c.vanlaerhoven@gmail.com}, R. Barnes$^{2,3}$ and R. Greenberg$^{1}$\\
$^{1}$Department of Planetary Sciences, University of Arizona, 1629 E University Blvd, Tucson, AZ, 85721, USA\\
$^{2}$Astronomy Department, University of Washington, Box 351580, Seattle, WA, 98195, USA \\
$^{3}$NASA Virtual Planetary Laboratory, USA}

\begin{document}

\date{in original form January 2014}

\pagerange{\pageref{firstpage}--\pageref{lastpage}} \pubyear{2014}

\maketitle

\label{firstpage}

\begin{abstract}
Earth-scale planets in the classical habitable zone (HZ) are more likely to be habitable if they possess active geophysics. Without a constant internal energy source, planets cool as they age, eventually terminating tectonic activity and rendering the planet sterile to life.  However, for planets orbiting low-mass stars, the presence of an outer companion could generate enough tidal heat in the HZ planet to prevent such cooling. The range of mass and orbital parameters for the companion that give adequate long-term heating of the inner HZ planet, while avoiding very early total desiccation, is probably substantial. We locate the ideal location for the outer of a pair of planets, under the assumption that the inner planet has the same incident flux as Earth, orbiting example stars: a generic late M dwarf ($T_{eff}=2670 K$) and the M9V/L0 dwarf DEN1048. Thus discoveries of Earth-scale planets in the HZ zone of old small stars should be followed by searches for outer companion planets that might be essential for current habitability.
\end{abstract}

\begin{keywords}
celestial mechanics - planetary systems - astrobiology
\end{keywords}

\section{Introduction}\label{intro}

In the search for habitable planets, Earth-scale objects are preferentially found very close to their primary stars because both major discovery techniques tend to favor finding them there. For a close-in planet, transits are less affected by inclination and stellar radial velocities are more sensitive to a planet's gravity. However, such orbits often put the planets closer to the star than the traditional ``habitable zone'' \citep{Kas93}, because the stellar radiation is too intense. Extreme examples are Kepler-10b \citep{Bat11}, Kepler 78b \citep{San13} and the four planets reported by Jackson et al. (2013). Therefore, the best chances to find a habitable planet may be near a low-mass star (e.g. an M or K star), where the habitable zone can have a small inner radius (\citealt{Kop13}, \citealt{Yan13}).

Calculating the dimensions of the habitable zone is a complex and uncertain process, even with its restricted, classical definition based on sustaining liquid water on the surface. Low mass stars differ from the Sun not only in the lower total radiation, but also in the spectrum, which is important because of the wavelength dependence of planetary albedo. For example, surface ice is very reflective in the visible light from Sun-like stars, but its albedo is low in the infrared, which is the peak emission from these low-mass stars (\citealt{Jos12}, \citealt{Shi13}). Similar issues affect consideration of the role of clouds \citep{Yan13}. 
Since M dwarf stars cover a very wide range of stellar temperatures, the distance of the habitable zone boundaries is not the same for all M dwarfs. In this paper we consider a planet that receives the same total incident flux as the Earth does ($1370 ~W/m^2$). For a star of mass 0.1 solar masses and luminosity $1.15\times10^{-3}$ times solar ($T_{eff}=2670$ K), this planet's semi-major axis is 0.034 AU. Despite the various uncertainties, for low-mass stars the habitable zone likely coincides with the region where Earth-scale planets are most readily discoverable.

On the other hand, given the long lifetimes of low-mass stars, a substantial fraction of them are very old, some more than twice as old as the Earth. Moreover, because the older ones are less active than younger ones, their planets may be more easily detected. Thus the most readily discoverable Earth-scale, habitable-zone planets are likely to be $\sim10$ Gyr old. 

Despite their location within a habitable zone, the extreme ages of such planets present a challenge to habitability because such planets are likely to have cooled internally. For example, terrestrial life depends on heat-driven plate tectonics to maintain the carbon cycle and to moderate the greenhouse effect. On an Earth-like body, long before reaching twice Earth's age, plate tectonics would probably have turned off as the planet cooled, primarily because solidification of the core would terminate the release of latent heat that drives mantle convection (e.g. \citealt{Sle00}, \citealt{vSu13}). While plate tectonics may not be essential for life on all habitable planets, an equivalent tectonic process to drive geochemical exchange between the interior and the atmosphere is a likely requirement. The necessary amount of internal heat for such activity is uncertain (even the mechanisms that govern the ontset and demise of terrestrial plate tectonics are still poorly understood and controversial), but it seems likely that a planet $\sim$10 Gyr would have cooled too much. A previously habitable planet, even if it remains in the classical habitable zone, might now be uninhabitable.

These considerations suggest that current habitability would require an additional internal heat source. One potential mechanism is heating by tidal friction. Tidal heating requires that the body be continually worked as the tidal potential deforms it. The planet must be either rotating non-synchronously as it orbits the star or its orbit must be eccentric.  However, tides tend to turn off the heating by synchronizing the rotation and circularizing the orbit. Life can be supported only if the despinning or circularization occur slowly enough that adequate heating is maintained on geologic timescales.

The spin-down is very rapid. Consider an M-type star with 0.1 solar mass (like the example above) orbited by an Earth-sized (mass and radius) rocky planet in the habitable zone at 0.034 AU where its ``instellation'' is about the same as the insolation at the Earth. Using a conventional formulation for the tidal spin-down timescale (e.g. \citealt{PeaXX}, or \citealt{Gla96}), and assuming that the tidal parameter $Q' \equiv (3/2)Q/k_2$ (where $k_2$ is the second-degree Love number) has a value 100,which is a typical estimate for a rocky planet, this timescale is a few thousand years, so the planet would have become synchronous essentially as it formed.

The eccentricity would also have been damped very early in such a planet's history. The damping rate (e.g. \citealt{Gol66}, \citealt{Kaula}, \citealt{VL13}) is

\begin{equation}\label{dedt}
\frac{de_p/dt}{e_p} = - \frac{63}{4} \sqrt{GM_s^3} \frac{R_p^5}{Q'_p m_p} a_p^{-13/2}.
\end{equation}

\noindent where $e_p$ is the eccentricity of the planet, $M_s$ is the mass of the star, $m_p$ and $R_p$ are the mass and radius of the planet, and $a_p$ is the planet's semi-major axis. The eccentricity thus damps on a timescale of $\sim$0.2 Gyr.  Equation \ref{dedt} accounts only for tides raised on the planet.  Tides raised on the star by the planet can also affect the eccentricity, but in this case that effect would be negligible due to the higher stellar $Q'$ \citep{VL13}. The eccentricity-damping timescale strongly depends on $a$. Because $a$ is also changed by tides, it is important to check whether that change can affect the eccentricity damping timescale \citep{JacXX}.  The semi-major axis changes at the rate:

\begin{equation}\label{dadt}
\frac{da_p/dt}{a_p} = e_p^2 \frac{de_p/dt}{e_p}
\end{equation}

\noindent Assuming that the initial eccentricity is not $\sim$1, the timescale for damping of the eccentricity of such a planet would remain well less than a Gyr. Hence tidal heating could not have been a significant factor in this planet's long-term geophysical evolution.

One way to maintain the orbital eccentricity, and hence the requisite tidal heating for active tectonics on a planet orbiting a very old star, would be by perturbations from an adjacent outer planet. 
Here we consider the secular perturbations between planets (e.g. \citealt{Bro61}, \citealt{Mur99}, \citealt{Bar06}, \citealt{VL13}), which we show below yield stable long-term tidal heating in non-resonant systems.

For a two-planet system, the secular interactions can be characterized by two eigenmodes, each with an amplitude determined by the initial conditions (initial eccentricities and pericenter longitudes) of the two planets (e.g. \citealt{VL12}). For each eigenmode, the eigenvector determines how the amplitude is shared between the two planets. The values of the orbital eccentricities oscillate under the influence of the two eigenmodes.

Generally, one eigenmode damps much faster than the other, with few exceptions \citep{VL13}. The damping time of the faster damping mode is typically fairly short, because it is comparable to the eccentricity damping that would have been experienced by the planet most prone to tidal effects (usually the inner planet) if the other planet did not exist (i.e. the rate given by Eq. \ref{dedt} above). Once that fast-damping mode has turned off, the planets' eccentricities stop oscillating and reach nearly fixed values, determined by the remaining, slow-damping, eigenmode.

For an outer planet to induce an acceptable amount of tidal heating in the habitable-zone planet, the eccentricity corresponding to the long-lived eigenmode must generate heat great enough to drive active tectonics on the inner planet, but not so much that it sterilizes it. The eigenmode must be so long-lived that it endures for $\sim$10 Gyr.  Equally important is that the fast-damping eigenmode, during its short early lifetime, never generated enough tidal heat long enough to desiccate the planet \citep{Bar13}.

Here we explore the constraints on the architectures of systems that could allow for the current habitability of an inner planet in the classical habitable zone. In Section \ref{setup} we demonstrate the wide range of parameters for which systems have very different damping rates for the two eigenmodes.  Then in Section \ref{ltinst}, we compute the expected tidal heating for these systems when only the slow-damping mode remains and compare this heating to relevant examples. In Section \ref{evol} we consider the variation of tidal heating over time as that mode damps, and we also examine the evolution during the early history while both modes were active. Finally, in Section \ref{concl}, we discuss the implications for habitability, showing that there is a range of plausible system parameters for which adequate tidal heating has been maintained over the long term, without overheating the planet at any point during its evolution.

\section{Relative damping timescales for two eigenmodes}\label{setup}

For a single planet in an unperturbed keplerian orbit, the eccentricity (as well as its orientation given by the longitude of pericenter $\varpi$) is a constant of integration of the two-body (planet plus star) problem. For a non-resonant multi-planet system, classical secular theory describes the quite different behavior.  This theory has been presented in detail and frequently reiterated throughout the literature of planetary dynamics (e.g. \citealt{Bro61}, \citealt{Mur99}, \citealt{Bar06}, \citealt{Wu02}, \citealt{VL13}). Here we summarize its salient features. Rather than staying constant, the eccentricities vary periodically, according to several eigenmodes, each with its own eigenfrequency. The number of eigenmodes equals the number of planets.

The solution for the behavior also includes eigenvectors, which describe how the strength of each eigenmode is distributed among the planets. The eigenfrequencies and eigenvectors are determined by the semi-major axes and masses of the planets. The amplitudes of the eigenmodes, rather than the planets' eccentricities, are the constants of integration. Those constant amplitudes depend on the initial eccentricities and $\varpi$ (longitude of pericenter) values.

Suppose only one eigenmode had a significant amplitude. In that case the planets' eccentricities would have fixed values proportional to the components of the eigenvector, i.e.

\begin{equation}\label{1modeecc}
e_p = E_m V_{mp}
\end{equation}

\noindent where $e_p$ is the eccentricity of any planet $p$, $E_m$ is the amplitude of the eigenmode $m$ (the only one with a non-negligible amplitude), and $V_{mp}$ is the $p$ component of the eigenvector of mode m. More generally, all the modes might be expected to have finite amplitudes, in which case the eccentricities would have values that vary according to the periods of the various eigenmodes.

Now consider the effect of tidal damping of one planet's eccentricity according to Equation \ref{dedt} (above). Rather than simply affect the orbit of that planet, the amplitudes of all the eigenmodes are damped through the secular interactions.  Each mode damps at its own rate, although they all damp with a longer time scale than given by Equation \ref{dedt}. The calculation of these rates for various multiplanet systems was demonstrated by \citet{VL12} for example.

For a two-planet system, such as those we are considering here, formulae for the damping rates of the amplitudes $E_m$ of the eigenmodes are given by \citealt{VL13} (Appendix). To confirm that one mode generally damps much faster than the other, consider the ratio of the damping rates:

\begin{equation}\label{D}
D \equiv \frac{\dot{E}_1 / E_1}{\dot{E}_2 / E_2} = \frac{-V_{11} V_{22}}{V_{12} V_{21}}
\end{equation}

\noindent (\citealt{VL13}, Eq. A52). This formula assumes that only the inner planet's eccentricity is directly damped by tides. Here we let mode 1 be the eigenmode that damps fastest as tides act on the inner planet and that the evolution of the planets' semi-major axes is negligible. As shown by \citep{VL13}, the fastest damping eigenmode also has the higher eigenfrequency.  Also, for that mode the components of the eigenvector have opposite signs; in other words, $-V_{11}/V_{12} > 0$, so if the second mode were very weak, the major-axes of the planets orbits would be anti-aligned with pericenters $180^o$ apart. Accordingly, the right-hand side of Equation \ref{D} is positive. All the $V$ components depend on the masses and semimajor axes of the two planets, and here we also include general relativistic effects on the precession of the apsides, which also contribute to the values of the $V$ components.

\begin{figure}
\includegraphics[scale=0.6]{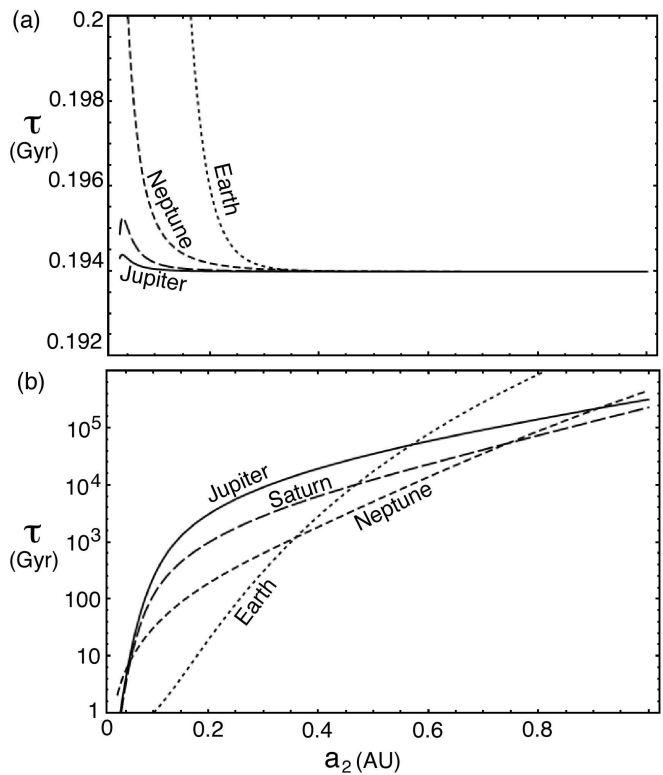}
\caption{The damping timescales for the two eigenmodes of the secular behavior as a function of the mass $m_2$ and semi-major axis $a_2$ of the outer planet. The inner planet has a mass equal to that of the Earth, and is located at 0.034 AU around a star of 0.1 solar masses. (a) The top panel shows the fast-damping mode (\#1); (b) The bottom panel shows the slow-damping mode (\#2).}
\label{timescales}
\end{figure}

We consider the nominal inner planet assumed in Section \ref{intro}:  an Earth-mass inner planet at semimajor axis $a_1$ = 0.034 AU with an outer companion, orbiting a 0.1 Solar-mass star. Equations A50 and A51 from \citep{VL13} give the damping timescales of each mode as a function of the V components. Figures \ref{timescales}a and b show these timescales for the fast-damping mode (\#1) and the slow-damping mode (\#2) as a function of the semimajor axis $a_2$ and mass $m_2$ of the outer planet.

The general trends of these functional dependences follow from the forms of the cited equations, but they can also be understood in terms of the coupling of each of the modes between the two planets. If the planets are very far apart, the orbits are nearly uncoupled. In that case, mode 1 describes the motion of the inner planet with little relation to the orbit of the outer one; and mode 2 describes the motion of the outer planet with little relation to the inner planet.  In this limit, the amplitude of mode 1 nearly equals the eccentricity of the inner planet and vice versa.  Thus, the damping timescale of mode 1 is equal to the damping timescale of $e_1$ as if there were no outer planet, which is the value 194 Myr shown for large $a_2$ in Figure \ref{timescales}a. For a given mass $m_2$, for smaller $a_2$, the mode is shared between the planets, so tides on the inner planet have less effect on mode 1 and the timescale is longer. Similarly, as shown in Figure \ref{timescales}b, if $a_2$ is large enough, the outer planet's orbit is independent of any damping of the inner planet, so mode 2 will have negligible damping (very long timescale). But as $a_2$ is decreased, mode 2 is shared more between the two planets and it is damped by inner planet tides.

Figure \ref{timescales} also shows the dependence of damping times on $m_2$. For mode 1 (Figure \ref{timescales}a) as the mass of $m_2$ decreases from that of Jupiter toward that of the Earth, mode 1 damps somewhat more slowly. The reason is that the inner planet shares more of its dominant mode with the outer planet, so inner planet tides take longer to damp the two planets.  However, as the outer planet gets even smaller (not shown in Figure \ref{timescales}a), eventually it would have no effect on the inner planet, so the latter would damp independently, i.e. mode 1 would damp on the 194 Myr timescale for all $a_2$ if $m_2$ were small enough.

The dependence on $m_2$ of the damping rate of mode 2 can be interpreted similarly. For example, Figure \ref{timescales}b shows that for $a_2$=0.07, as $m_2$ varies from Earth's mass to Jupiter's, the damping rate decreases, because for an extremely massive outer planet, $m_1$ has no effect on it. Thus mode 1 acts entirely on the outer planet and inner-planet tides have little effect on damping that mode.  For larger values of $a_2$ the dependence on $m_2$ gets more complicated: for very small $m_2$, the planets become uncoupled, so again inner-planet tides would only weakly damp mode 1. However, that effect only comes into play for cases where the damping times would far exceed the age of the universe.

\begin{figure}
\includegraphics[scale=0.6]{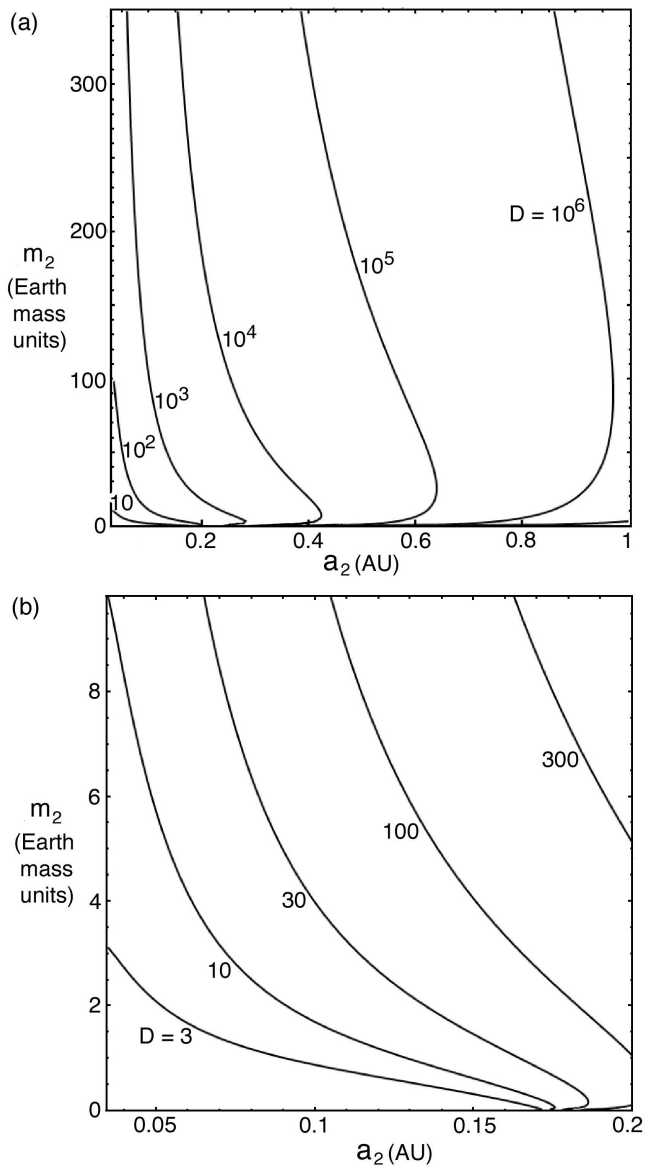}
\caption{a) The ratio $D$ of the damping rates of the two eigenmodes as a function of the mass and semi-major axis of the outer planet. $D$ is generally $\gg 1$. b) Enlargement of the portion for small outer planet mass $m_2$ and semi-major axis $a_2$. Only for the smallest values do the two modes damp at comparable rates.}
\label{Dplot}
\end{figure}

Over a wide range of masses and orbits for the outer planet, mode 1 damps much faster than mode 2, i.e. the ratio $D \gg 1$, as shown in Figure \ref{Dplot}.  Only when the mass of the outer planet is small and its semi-major axis is quite close to that of the inner planet, as shown in the expanded plot of this portion of the parameter space, do the two damping timescales become similar. Note that over the course of long-term tidal evolution these timescales would change slightly because of the change in the values of the $V$ components as $a_1$ changes, but the conclusion is not affected: Mode 1 damps away very early in the evolution.

Thus, we can separate the evolution of systems into the short-term part, where both modes contribute significantly to the planets' eccentricities, and the long-term part, where only mode 2 contributes to the planets' eccentricities. Next in Section \ref{ltinst}, we consider the latter stage to determine what parameters produce a long-term tidal heating rate adequate for active tectonics and habitability. Later in Section \ref{evolst} we consider additional constraints based on the stage where both modes contributed heating. Habitability will require that the inner planet not be over heated during that stage as well.

\section{Tidal Heating driven by the long-lived secular eigenmode}\label{ltinst}

The previous section shows that most systems of interest quickly evolve to single eigenmode behavior.  At that point, the eccentricities no longer oscillate and their values decrease only slowly.

Next we consider the tidal heating of the inner Earth-like planet in such a system.  We assume that it orbits a low-mass star at a distance where its ``instellation'' matches the insolation of the Earth at 1 AU from the Sun. In other words, it lies well within the classically defined habitable zone. Specifically (as above) we assume a star of 0.1 Solar masses with luminosity of $1.15\times10^{-3}$ times solar. This yeilds $a_1 = 0.034 AU$. The planet is assumed (as above) to to rocky, having the Earth's mass and radius. We assume the tidal dissipation parameter $Q'_1$ is 100. This value lies between the current low $Q'$ value of $\sim50$ for the Earth, which reflects to some degree the current arrangement of the continents, and the seismic $Q'$ for the mantle of $\sim1000$. The objective is to identify the range of the other system parameters (outer planet mass and semi-major axis, and the secular eigenmode amplitude) for which the heating would be adequate to maintain the active geophysics needed for life.

The required amount of heat is poorly known. The mechanisms for initiation or maintenance of plate tectonics on an Earth-like planet are so complex (e.g. \citealt{Sle00}) that a required minimum heat flux can only be estimated.  To a first approximation, an amount equivalent to the current surface heat flux from the Earth's interior, $\sim0.08 ~W/m^2$, would be reasonable.

As a point of reference, on Mars plate tectonics may have shut down when the internal heat flux reached $0.04 ~W/m^2$ \citep{Wil97}, so that value might be taken as a crude minimum.  Within the Earth, about 1/3 of the heat comes from phase transitions in the core and the remainder from radiogenic heating in the mantle. As the Earth cools, in another 4 Gyr the core will solidified (e.g. \citealt{vSu13}) and radiogenic heating will be halved, reducing the heat flux to $<0.03 ~W/m^2$. (We are making some assumptions here about the radiogenic inventory of the Earth. See \citet{Ara05}, \citet{Dye12}.) From the Mars comparison, tidal heating would need to provide at least an additional $0.01 ~W/m^2$ to keep the Earth's heating above the critical level, but tidal heating may not be deep enough to drive deep mantle convection. Several times that much heat might be required to drive the required tectonic activity.

At the opposite extreme, too much tidal heating could lead to excessive volcanism, like that displayed by Io, with consequent rapid resurfacing. 
Io's surface heat flux is about $2 W/m^2$, but the actual upper limit to acceptable heating for life could be much lower than that value, unless the planet has far more water than Earth. 

Here we calculate the conditions under which tides can deliver amounts of heat ranging over two orders of magnitude from 0.02 to 2 $W/m^2$. 

The tidal heating is given by

\begin{equation}\label{heat}
h_1 = \frac{63}{16 \pi} \frac{(G M_s)^{3/2} M_s R_1^3}{Q'_1} e_1^2 a_1^{-15/2}
\end{equation}

\noindent where $M_s$ is the mass of the star, $R_1$, $e_1$ and $a_1$ are the radius, eccentricity, and semi-major axis of the inner planet, and $Q'_1$ is the tidal dissipation factor for the inner planet. The eccentricity of the inner planet depends on the amplitude of the remaining eigenmode ($E_2$), and how that amplitude is shared between the planets as given by the eigenvector (Eqn \ref{1modeecc}). Note that whatever the choice of $E_2$, both eccentricities are smaller than $E_2$ and they satisfy $e_1^2 + e_2^2 = E_2^2$.

Given the fixed parameter choices above for $M_s$, $R_1$, $Q'_1$ and $a_1$, the inner planet's heating rate $h_1$ depends only on $e_1$.  In turn, $e_1$ is a function of three remaining parameters of the system: the outer planet's semi-major axis and mass ($a_2$ and $m_2$) and the amplitude of the long-lived eigenmode ($E_2$). The planets' masses and semi-major axes yield the eigenvector, and the eigenvector combined with $E_2$ yields the eccentricities of both planets.

\begin{figure}
\includegraphics[scale=0.6]{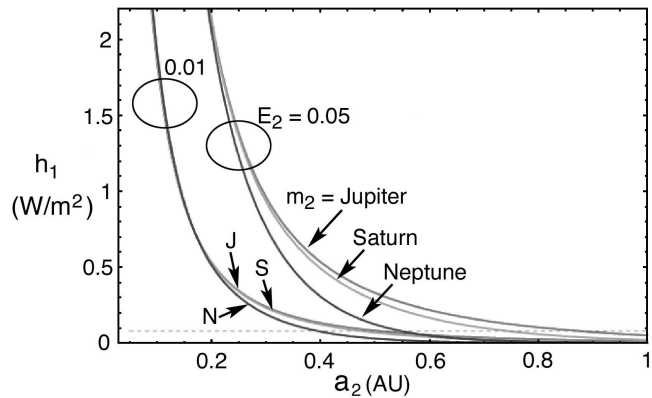}
\caption{The tidal heating rate experienced by the inner planet with a Jupiter mass, Saturn mass, or Neptune mass outer planet for example amplitudes of mode 2 of $E_2 = 0.01$ and $0.05$. For each example $E_2$ the values of $h_1$ are tightly grouped, despite the large mass differences of the outer planets.}
\label{h1vsa2}
\end{figure}

\begin{figure}
\includegraphics[scale=0.6]{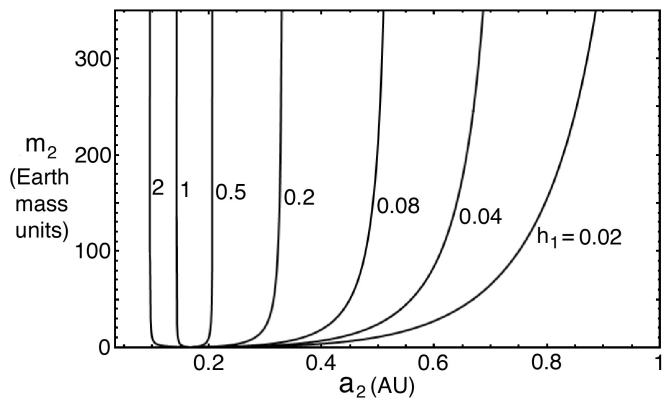}
\caption{The tidal heating rate of the inner planet ($W/m^2$) as a function of the outer planet's mass $m_2$ (in $M_{Earth}$) and semi-major axis $a_2$ (in AU) assuming the amplitude of the second eigenmode is $E_2=0.05$. For large $m_2$ the lines of constant $h_1$ are nearly vertical, indicating that the tidal heating is nearly independent of $m_2$. Only when the planets are very well separated and/or $m_2$ is less than a couple tens of Earth masses does the tidal heating rate depend strongly on $m_2$.}
\label{h1vsm2a2}
\end{figure}

\begin{figure*}
\includegraphics[width=\textwidth]{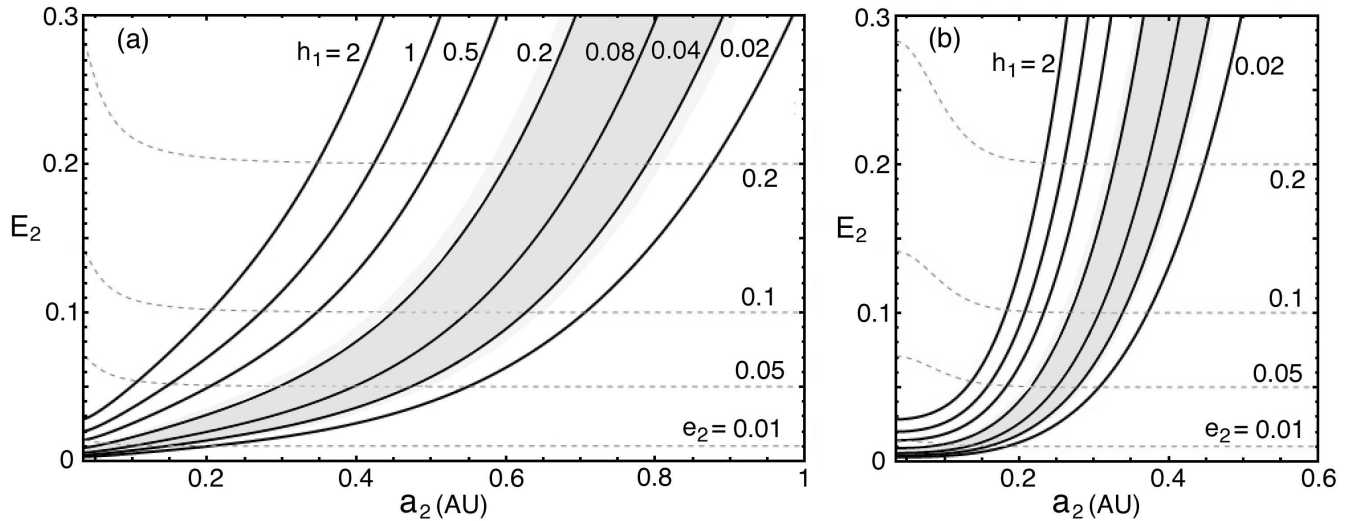}
\caption{The tidal heating rate of the inner planet as a function of the amplitude of the eigenmode $E_2$ outer planet's semi-major axis $a_2$ (in AU), for an outer planet mass $m_2$ equal to the mass of a) Neptune and b) Earth.}
\label{h1vsE2a2}
\end{figure*}

Thus the inner planet's heating rate $h_1$ depends on the three parameters $m_2$, $a_2$ and $E_2$, acting through the secular perturbations. We can look at this functional dependence in several ways. Figure \ref{h1vsa2} shows $h_1$ as a function of $a_2$ for two representative values of $E_2$ and three values of $m_2$ (the masses of Jupiter, Saturn, and Neptune, respectively). Figure \ref{h1vsm2a2} shows $h_1$ as a function of $a_2$ and $m_2$ for a specific choice of $E_2=0.05$. Figure \ref{h1vsE2a2} shows how $h_1$ depends on $a_2$ and $E_2$ if $m_2$ equals the mass of Neptune or the mass of Earth.

Figure \ref{h1vsa2} shows that for the given $E_2$ values, the mass of the outer planet only weakly affects the inner planet's tidal heating rate. Figure \ref{h1vsm2a2} also shows that $h_1$ depends only weakly on $m_2$ unless $m_2$ is less than several Earth masses. As shown in Figure \ref{h1vsm2a2}, for $E_2=0.05$, for $m_2 > 20$ Earth masses, a value of $a_2$ near 0.5 AU will given the requisite internal heating. A less massive outer planet would need to be closer in, for example near 0.25 AU if $m_2 =1$ Earth mass.

In addition to showing the heating rate $h_1$ as a function of $E_2$ and $a_2$, Figure \ref{h1vsE2a2} relates $h_1$ to the eccentricity $e_2$ of the outer planet in the following way. From Equation \ref{heat}, any given heating rate corresponds to a particular value of $e_1$, so along any constant-$h_1$ contour the variation of $E_2$ corresponds to the variation of the eccentricity of the outer planet, which is indicated by the dashed contour lines of constant $e_2$.  Thus these plots show how the tidal heating varies with $e_2$. For example, according to Figure \ref{h1vsE2a2}a, to provide tidal heating of the habitable-zone planet of $0.08 W/m^2$, a Neptune mass at 0.55 AU would need an eccentricity of 0.1. If it were at 0.4 AU, it would need an eccentricity of 0.05. Similarly, for an outer planet of one Earth mass, the required eccentricity would decrease with $a_2$ as shown in Figure \ref{h1vsE2a2}b.

As a reference, Figure \ref{h1vsE2a2} includes shading over the portion of outer-planet parameter space that yields heating within about a factor of two of the internal heat flux of the Earth, which seems a plausible range for habitability. The shading has a deliberately fuzzy boundary to reflect the uncertainty in the actual limits on heat flux.  As understanding of the geophysical constraints on life, and thus the heat-flux limits, improves, one might want to adjust the range of acceptable outer-planet parameters according to the other contours in Figure \ref{h1vsE2a2}. Note that while we explicitly display results only for heating rates between 0.02 and 2 $W/m^2$, the results could be readily extrapolated because the contours are spaced approximately logarithmically (e.g. the spacing between contours for $h_1 = 1$ and $2 ~W/m^2$ is similar to that between 0.04 and 0.08 $W/m^2$).

A hard lower limit to $a_2$ is given by stability constraints. For stability, the planets should get no closer to one another than about 3 mutual Hill radii from each other (e.g. \citealt{Gla93}, \citealt{Smi09}). If $m_2 = 1$ Neptune mass, the lower limit to $a_2$ is 0.04 AU; If the outer planet's mass equals one Earth mass (the same as the inner planet), its lower limit would be 0.037 AU. For larger values of $a_2$, the required eccentricity increases, but never so much that it would bring pericenter so close to the inner planet as to destabilize the system.

The qualitative dependence of $h_1$ on $m_2$, $a_2$ and $E_2$ depends on the basic mechanism of secular interaction. Figures \ref{h1vsa2}, \ref{h1vsm2a2} and \ref{h1vsE2a2} show that heating of the inner planet decreases as $a_2$ increases as expected, because with the greater separation the planets become increasingly decoupled so that less of the eigenmode amplitude $E_2$ is shared with the inner planet. This decrease in the portion of $E_2$ that goes into the eccentricity (and heating) of the inner planet is also apparent from the dashed lines in Figure \ref{h1vsE2a2}, which show that, as $a_2$ increases, $e_2$ and $E_2$ converge toward the same value.

Similarly, for a given $a_2$, as $m_2$ increases the tidal heating of the inner planet increases (e.g. Figure \ref{h1vsm2a2}), because the planets' orbits because more tightly coupled.  However, as the value of $m_2$ becomes large, the eigenvector of the long-lived eigenmode becomes nearly independent of $m_2$. 

In summary, this section shows that reasonable masses and orbits for a second planet can readily drive the eccentricity of an Earth-scale habitable-zone planet, and hence its tidal heating, to a degree that might plausibly maintain plate tectonics. The next section addresses the evolution of such tidal heating over time.

\section{Evolution}\label{evol}

According to the results in Section \ref{setup}, by sharing the secular eigenmodes, an outer companion planet may be able to promote enough tidal heating in an inner planet to meet the requirements for habitability. However, over time the tides also act to change the orbits, so we need to consider over how much of the inner planet's lifetime the tidal heating is adequate to maintain habitability, and if or when it may be so great as to sterilize the planet. Tides raised on the planet tend to decrease the eigenmode amplitudes (e.g. \citealt{VL13}), which decreases $h_1$ according to Equation \ref{heat}. Also, $a_1$ decreases due to tides on the planet and the star, which increases $h_1$. (The planet we are considering here is inside the co-rotation distance from the star; otherwise tides on the star would tend to push the planet's orbit outwards). Thus, to have a stable $h_1$, either there should be very little evolution of either the eigenmode amplitudes or $a_1$, or their evolution should counteract each other's influence on $h_1$.

If the semi-major axis evolves significantly a planet could exit the classical habitable zone and receive too much instellation from its host star \citep{BarXX}. Such migration would be minimal for a single planet, because its eccentricity would damp, slowing the semi-major axis evolution. In our multi-planet systems $e_1$ is maintained over the long duration of the slow-damping eigenmode, which increases the total change in semi-major axis. The corresponding increase in instellation, along with tidal heating, is another time variable factor in governing habitability.

In order to explore the range of possible evolutionary scenarios, we have considered the orbital evolution and corresponding heating rates for a wide range of system parameters for the same hypothetical planet considered above, that is the Earth-mass inner planet initially at 0.034 AU around the star of 0.1 solar mass. In our simulations we follow the procedure outlined in \citet{VL13} to evolve the secular eigenmode amplitudes and the planetary semi-major axes. We include tides on both planets and tides raised on the star (with $Q_{star}=10^6$) in our simulations. Tides do cause the semi-major axis of the outer planet to change, but we find that effect is minimal. 

\begin{figure}
\includegraphics[scale=0.6]{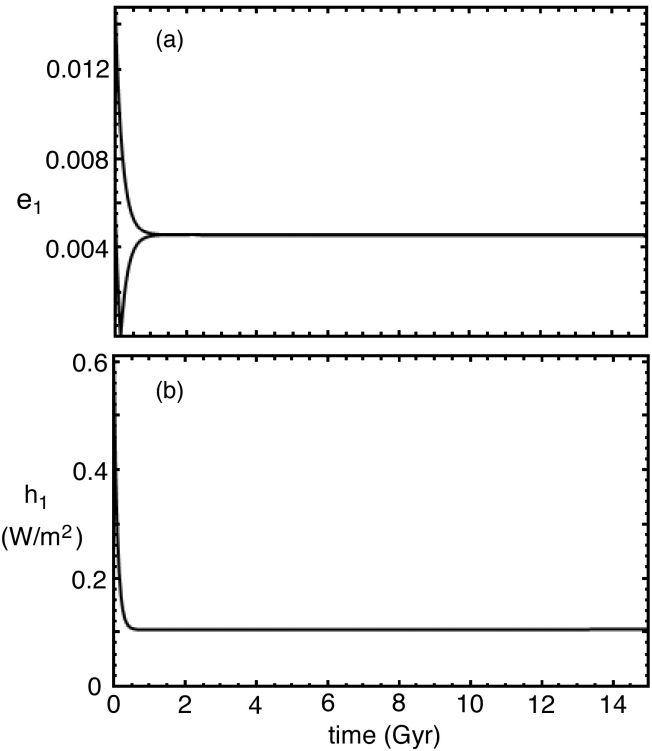}
\caption{An example of the evolution of the inner planet's eccentricity $e_1$ and tidal heating rate $h_1$ ($W/m^2$). a) Maximum and mimimum eccentricity versus time. b) Average tidal heating rate for the inner planet $h_1$ versus time. }
\label{typicalevol}
\end{figure}

Figure \ref{typicalevol} shows an example of the evolution for a system with a Neptune-mass outer planet initially with $a_2=0.45$ AU; initial $E_1 = 0.01$, and $E_2 = 0.05$. In Figure \ref{typicalevol}a, the two curves represent the range of variation of $e_1$ during the secular oscillation of its value. Initially $e_1$ oscillates about a value determined by mode 1 with an amplitude given by mode 2.  After 200 Myr, the fast damping mode 1 has been reduced enough that $e_1$'s oscillation reaches zero.  At this point in the evolution, the pericenters transition from circulation to libration (see e.g. \citealt{Bar06}, \citealt{Chi02}). Thereafter, $e_1$ oscillates about the value given by mode 2 with an amplitude given by mode 1. Then mode 1 largely damps away by about 1 Gyr. The eccentricity of the inner planet is then set only by mode 2. By this time, $e_1$ is so small that there is very little tidal effect on $a_1$. The amplitude of the slow damping mode 2 decreases very little over 15 Gyr, and $a_1$ remains nearly constant. Thus, this inner planet receives a steady tidal heat flux and instellation after about the first Gyr.

In order to determine what part of the parameter space yields long-term tidal heating adequate to maintain active geophysics (and hence habitability), we first require that it be adequate during the long period after mode 1 has damped. The range of systems that satisfy that criterion is considered in Section \ref{evollt}. Then, in Section \ref{evolst}, we eliminate from that set those systems that would have suffered extreme sterilizing heat during the early phase while both modes have significant amplitudes. The remaining range of systems could have allowed an Earth-scale planet in the habitable zone to have had the potential for supporting life, and to have then remained habitable until these low-mass stars reached their current great ages.

\subsection{Long Term heating rates from the slow-damping eigenmode}\label{evollt}

For the range of system parameters considered in Section \ref{ltinst} (c.f. Figure \ref{h1vsE2a2}), the amount of tidal evolution over 15 Gyr is shown in Tables \ref{E2h1m2N} and \ref{E2h1m2E}, for outer-planet masses equal to Neptune's and Earth's respectively. These two tables span the same parameters space ($E_2$ and $a_2$) as Figure \ref{h1vsE2a2}a and b respectively. Over the portion of this space in Figure \ref{h1vsE2a2} where tidal heating $h_1$ is appropriate for life support, Tables \ref{E2h1m2N} and \ref{E2h1m2E} show that the $h_1$ changes by $<1$\% during this long period of time. In fact, the change only exceeds 1\% where the secular eigenmode is very strong ($E_2\sim0.3$), $a_2<0.3$, and $m_2=$Neptune's mass. However, Figure \ref{h1vsE2a2}a shows that the inner planet would receive $\gg 2 W/m^2$ under such extreme circumstances, putting its habitability in jeopardy. 

So over this full range of parameter space, if the system parameters ($a_2$ and $E_2$) yeild a long-term tidal heating rate that is consistent with supporting conditions for life, we know that the system will not evolve away from this state for $t>15$ Gyr.

For small values of $a_2$, Tables \ref{E2h1m2N} and \ref{E2h1m2E} show that the eccentricity of the inner planet can damp considerably, even as $h_1$ remains nearly constant.  This result includes a range of cases with $h_1$ values in the life-supporting range. How can $h_1$ remain constant as $e_1$ decreases? The answer is that the decrease in semi-major axis $a_1$ closely balances the decrease in $e_1$ in those cases.

The fact that $a_1$ changes over time raises the question of whether the planet is evolving out of the classical habitable zone to where the instellation is too great to allow liquid water at the surface. Tables \ref{instellm2N} and \ref{instellm2E} show how much the instellation changes over the 15 Gyr period for the evolution cases shown in Tables \ref{E2h1m2N} and \ref{E2h1m2E}. We have assumed that the star has a constant luminosity, which is reasonable for middle-aged to old M stars.

According to \citet{Yan13}, an Earth-like planet would not be habitable if the instellation were greater than about twice that currently received by the Earth. In all the cases discussed in this section, the inner planet started with $a_1=0.034$ AU, the value at which such the instellation would match Earth's. As shown in Table 2, the only conditions under which the instellation increases by more than 100\% (i.e. doubling) are for the Neptune-mass outer planet with $a_2\sim0.1$ and $E_2>0.15$, in which case $h_1$ would already be so great (outside the gray zone in Figure \ref{h1vsE2a2}) as to preclude habitability. (Only if we were considering lifetimes of many tens of Gyr would this effect be relevant to the gray zone in Figure \ref{h1vsE2a2}.) Otherwise, the instellation does not change enough to drive the planet out of the classical habitable zone.

We conclude that for those systems where tidal heating $h_1$ is of a magnitude likely to be consistent with habitability at the time that the first eigenmode has damped away, we expect that it will endure for many billions of years after radiogenic and core heat have been depleted, allowing maintenance of tectonic activity.

\begin{table*}
\caption{The percentage change in $h_1$ and $E_2$ over 15 Gyr when $m_2 =$ 1 Neptune mass.}
\begin{tabular}{c|cc|cc|cc|cc|cc}
\hline\noalign{\smallskip}
 & \multicolumn{10}{|c}{$a_{2}$ (AU)} \\
$E_2$ & \multicolumn{2}{c|}{0.1} & \multicolumn{2}{c|}{0.3} & \multicolumn{2}{c|}{0.5} & \multicolumn{2}{c|}{0.7} & \multicolumn{2}{c}{0.9}  \\
\hhline{~--||--||--||--||--||}\noalign{\smallskip}
 & $\Delta h_1$ & $\Delta E_2$ & $\Delta h_1$ & $\Delta E_2$ & $\Delta h_1$ & $\Delta E_2$ & $\Delta h_1$ & $\Delta E_2$ & $\Delta h_1$ & $\Delta E_2$ \\
\noalign{\smallskip}\hline\noalign{\smallskip}
0.30 & 232\% & -96\%		& 34.0\% & -16\% 	& 0.985\% & -2.7\% 	& 0.280\% & -0.9\% 	& 0.104\% & -0.3\% \\
0.25 & -0.39\% & -99\%		& 1.77\% & -11\% 	& 0.257\% & -2.6\% 	& 0.085\% & -0.8\% 	& 0.033\% & -0.3\% \\
0.20 & 0.17\% & -80\%		& 0.204\% & -9.3\%	& 0.054\% & -2.4\% 	& 0.020\% & -0.8\% 	& $<$0.01\% & -0.3\% \\
0.15 & $<$0.01\% & -52\%	& 0.020\% & -8.1\%	& $<$0.01\% & -2.3\%	& $<$0.01\% & -0.8\% 	& $<$0.01\% & -0.3\% \\
0.10 & $<$0.01\% & -41\%	& $<$0.01\% & -7.4\%	& $<$0.01\% & -2.2\%	& $<$0.01\% & -0.8\% 	& $<$0.01\% & -0.3\% \\
0.05 & $<$0.01\% & -36\%	& $<$0.01\% & -7.0\%	& $<$0.01\% & -2.2\%	& $<$0.01\% & -0.8\% 	& $<$0.01\% & -0.3\% \\
0.001 & $<$0.01\% & -35\%	& $<$0.01\% & -6.9\%	& $<$0.01\% & -2.2\%	& $<$0.01\% & -0.8\% 	& $<$0.01\% & -0.3\% \\
\noalign{\smallskip}\hline
\end{tabular}
\label{E2h1m2N}
\end{table*}

\begin{table*}
\caption{The percentage change in $h_1$ and $E_2$ over 15 Gyr when $m_2 =$ 1 Earth mass.}
\begin{tabular}{c|cc|cc|cc|cc|cc}
\hline\noalign{\smallskip}
 & \multicolumn{10}{|c}{$a_{2}$ (AU)} \\
$E_2$ & \multicolumn{2}{c|}{0.1} & \multicolumn{2}{c|}{0.3} & \multicolumn{2}{c|}{0.5} & \multicolumn{2}{c|}{0.7} & \multicolumn{2}{c}{0.9}  \\
\hhline{~--||--||--||--||--||}\noalign{\smallskip}
 & $\Delta h_1$ & $\Delta E_2$ & $\Delta h_1$ & $\Delta E_2$ & $\Delta h_1$ & $\Delta E_2$ & $\Delta h_1$ & $\Delta E_2$ & $\Delta h_1$ & $\Delta E_2$ \\
\noalign{\smallskip}\hline\noalign{\smallskip}
0.30 & -0.810\% & -99\% 	& -0.00514\% & -5.1\% 	& $<$0.01\% & -0.10\% & $<$0.01\% & -0.06\% & $<$0.01\% & $<$0.01\% \\
0.25 & -0.391\% & -99\% 	& -0.0140\% & -5.1\% 	& $<$0.01\% & -0.10\% & $<$0.01\% & -0.06\% & $<$0.01\% & $<$0.01\% \\	
0.20 & -0.160\% & -99\% 	& $<$0.01\% & -5.1\% 	& $<$0.01\% & -0.10\% & $<$0.01\% & -0.06\% & $<$0.01\% & $<$0.01\% \\
0.15 & -0.0506\% & -99\% 	& $<$0.01\% & -5.1\% 	& $<$0.01\% & -0.10\% & $<$0.01\% & -0.06\% & $<$0.01\% & $<$0.01\% \\
0.10 & -0.0100\% & -99\% 	& $<$0.01\% & -5.1\% 	& $<$0.01\% & -0.10\% & $<$0.01\% & -0.06\% & $<$0.01\% & $<$0.01\% \\
0.05 & $<$0.01\% & -99\% & $<$0.01\% & -5.1\% & $<$0.01\% & -0.10\% & $<$0.01\% & -0.06\% & $<$0.01\% & $<$0.01\% \\
0.001 & $<$0.01\% &  -99\% & $<$0.01\% & -5.1\% & $<$0.01\% & -0.10\% & $<$0.01\% & -0.06\% & $<$0.01\% & $<$0.01\% \\
\noalign{\smallskip}\hline
\end{tabular}
\label{E2h1m2E}
\end{table*}

\begin{table*}
\caption{The percentage change in the instellation of the inner planet over 15 Gyr when $m_2 =$ 1 Neptune mass.}
\begin{tabular}{c|ccccc}
\hline\noalign{\smallskip}
 & \multicolumn{5}{c}{$a_2$ (AU)} \\
$E_2$ & 0.1 & 0.3 & 0.5 & 0.7 & 0.9  \\
\noalign{\smallskip}\hline\noalign{\smallskip}
0.30 & 2350\% & 199.\% & 25.5\% & 8.72\% & 3.44\% \\
0.25 & 582.\% & 68.4\% & 16.0\% & 5.89\% & 2.37\% \\
0.20 & 189.\% & 31.2\% & 9.50\% & 3.68\% & 1.50\% \\
0.15 & 54.9\% & 14.4\% & 5.06\% & 2.03\% & 0.842\% \\
0.10 & 17.3\% & 5.7\% & 2.16\% & 0.893\% & 0.373\% \\
0.05 & 3.68\% & 1.34\% & 0.529\% & 0.222\% & 0.093\% \\
0.001 & $<$0.01\% & $<$0.01\% & $<$0.01\% & $<$0.01\% & $<$0.01\% \\
\noalign{\smallskip}\hline
\end{tabular}
\label{instellm2N}
\end{table*}

\begin{table*}
\caption{The percentage change in the instellation of the inner planet over 15 Gyr when $m_2 =$ 1 Earth mass.}
\begin{tabular}{c|ccccc}
\hline\noalign{\smallskip}
 & \multicolumn{5}{c}{$a_2$ (AU)} \\
$E_2$ & 0.1 & 0.3 & 0.5 & 0.7 & 0.9  \\
\noalign{\smallskip}\hline\noalign{\smallskip}
0.30 & 14.4\% & 2.64\% & 0.0654\% & $<$0.01\% & $<$0.01\% \\
0.25 & 9.76\% & 1.83\% & 0.0455\% & $<$0.01\% & $<$0.01\% \\
0.20 & 6.12\% & 1.17\% & 0.0292\% & $<$0.01\% & $<$0.01\% \\
0.15 & 3.39\% & 0.659\% & 0.0165\% & $<$0.01\% & $<$0.01\% \\
0.10 & 1.49\% & 0.293\% & $<$0.01\% & $<$0.01\% & $<$0.01\% \\
0.05 & 0.370\% & 0.0733\% & $<$0.01\% & $<$0.01\% & $<$0.01\% \\
0.001 & $<$0.01\% & $<$0.01\% & $<$0.01\% & $<$0.01\% & $<$0.01\% \\
\noalign{\smallskip}\hline
\end{tabular}
\label{instellm2E}
\end{table*}

\subsection{Early short-term heating by the fast damping mode}\label{evolst}

The range of masses, locations and eccentricities of the outer planet that can give the required tidal heating of the inner planet was identified in Section \ref{ltinst}. We have also seen (Section \ref{evollt}) that the heating rate remains stable after the short-lived eigenmode (mode 1) has damped away. However, earlier on, immediately after formation of the system when both modes are active, the inner planet would have experienced a higher level of tidal heating. Therefore, we need to determine the upper limits to the initial mode 1 amplitude that can ensure that the planet is not over heated and desiccated early in its history.

\begin{figure}
\includegraphics[scale=0.6]{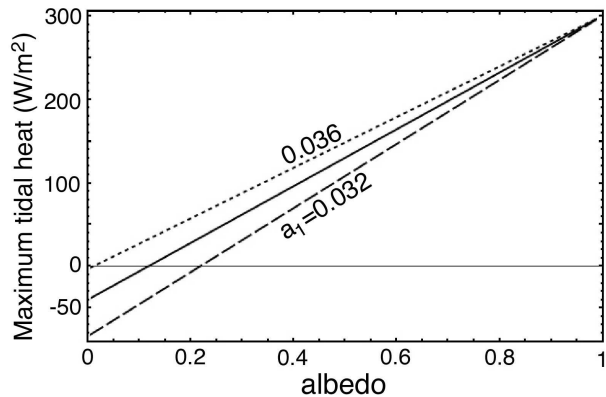}
\caption{The amount of heat that tidal heating can provide without driving the inner planet into a runaway greenhouse, as a function of the planet's albedo. The solid line is for aur nominal Earth-like planet at $a_1=0.034$ AU. The maximum amount of surface flux a planet can experience without triggering dessication is $\sim300 ~W/m^2$. Depending on the planet's albedo, absorbed energy from the star takes up most of this energy budget. Tidal heat can provide a maximum of $300 ~W/m^2$ minus the contribution from absorbed starlight without risking dessication. A negative value of maximum tidal heat indicates that for that albedo energy from the star alone results in a thermal surface flux of $> 300 W/m^2$.}
\label{critheat}
\end{figure}

An approximate requirement for desiccation of an Earth-like planet, such as the one we are considering, would be that the surface heat flux must exceed $\sim300 W/m^2$ for $\sim10^8$ yr \citep{Bar13}. If the planet had an initial excess abundance of water, the desiccation would take even longer, but to be conservative we assume that $10^8$ yr of such heating would sterilize the planet. The Earth-sized planet under consideration here receives the same instellation as Earth does, which is an incident flux of 1370 $W/m^2$. The surface flux is equal to the incident flux reduced by a factor $(1- albedo)/4$. For example, with the Earth's albedo of 0.3, the surface receives 240 $W/m^2$. For our hypothetical planet, the effective albedo could have a wide range of values, depending on cloud cover, surface materials, etc. As discussed in Section \ref{ltinst}, the contribution from radiogenic heat and from the latent heat released from the core is well below 1 $W/m^2$.  Thus the upper limit for acceptable tidal heating is approximately 60 $W/m^2$ for the Earth-like planet under consideration at 0.34 AU from the low-mass star.  This value depends on the albedo and on the distance from the star as shown in Figure \ref{critheat}. For example, if the albedo were about 0.15, only about 10 $W/m^2$ of tidal heat would be adequate for desiccation.

The tidal heating is proportional to $e_1^2$ (Equation \ref{heat}), and on the average over a secular cycle we have

\begin{equation}\label{e1}
e_1^2 = (E_1 V_{11})^2 + (E_2 V_{21})^2
\end{equation}

\noindent where the two terms on the right hand side are due to the two secular eigenmodes, respectively. Recall that $E_m$ represents the amplitude of the mode $m$ and $V_{mp}$ is a component of the eigenvector that shows how much of the amplitude affects the eccentricity of planet $p$. Equation \ref{e1} shows that the tidal heating is the sum of contributions from the two eigenmodes. 

As discussed in Section \ref{ltinst}, in the long term (after mode 1 has died) mode 2 cannot yield more than 2 $W/m^2$ or else the planet would become Io-like. This level of heating is negligible compared to the $300 W/m^2$ limit for dessication. Thus the upper limit to the acceptable heating by mode 1 alone is essentially the same (within $\lesssim 2 W/m^2$) as the upper limit for the total tidal heating shown in Figure \ref{critheat}. Therefore we need only consider the first term in Eqn \ref{e1} for evaluating the conditions for which the desiccation is unlikely.

Evaluation of $V_{11}$ shows that over nearly the full range of values of $m_2$ and $a_2$ for which long-term habitability of $m_1$ is possible (the zones marked in gray in Figure \ref{h1vsE2a2}), $V_{11}$ is nearly unity. (More precisely, for $m_2 =$ Neptune's mass, $V_{11}$ is within 1\% of unity for all $a_2$ down to its stability limit; for $m_2 =$ Earth's mass, $V_{11}$ is within 10\% of unity for all $a_2>0.08$ and within 25\% of unity even for $a_2$ at its stability limit of 0.037 AU.) Thus, $e_1 \simeq E_1$, which yields the heating rate via Eqn \ref{heat}. 

\begin{figure}
\includegraphics[scale=0.6]{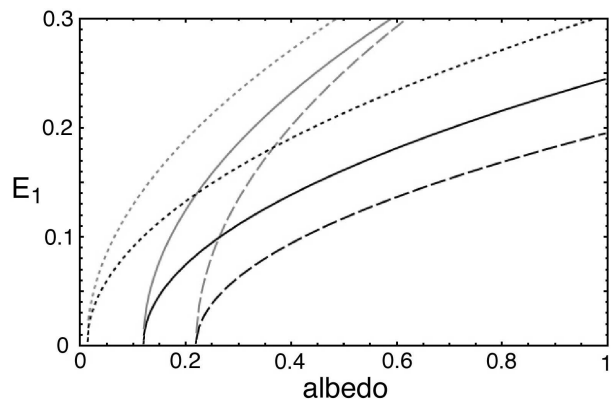}
\caption{Black curves: The value of $E_1$ that would result in the inner planet experiencing the critical flux for dessication (300 $W/m^2$), as a function of planetary albedo. Dashed, solid, and dotted lines correspond to $a_1=$ 0.032 AU, 0.034 AU, and 0.036 AU, respectively. Grey curves: The value of initial $E_1$ for which the planet would experience desiccating flux for $10^8$ years. If the initial $E_1$ lies below the grey line the planet will not experience high surface flux long enough to achieve dessication.}
\label{critE1}
\end{figure}

The solid black curve in Figure \ref{critE1} shows the values of $E_1$ that correspond to the tidal heat limit shown in Figure \ref{critheat}. For any value of $E_1$ above the curve, the total surface heat would exceed 300 $W/m^2$. However, because eigenmode 1 damps quickly, even if $E_1$ exceeds the critical value, it may not remain so for the full $10^8$ yr required for desiccation. The solid gray curve shows the initial value of $E_1$ that would be required to maintain the surface heat above 300 $W/m^2$ for the full $10^8$ yr.

These results show that within the range of acceptable systems from the point of view of long term heating (the zone marked gray in Figure \ref{h1vsE2a2}), systems in part of that range may have been prone to early desiccation, while for others early desiccation is less likely. 

For example, consider a system with an Earth mass outer planet at $a_2 = 0.5$ AU with $e_2 \simeq E_2=0.08$.  Since this system's parameters lie within the gray zone of the left panel in Figure \ref{h1vsE2a2}, once mode 1 damped away, this system would have generated adequate tidal heat to maintain tectonic activity essentially forever. Assuming that the initial amplitude $E_1$ of mode 1 was comparable to $E_2$, and that the inner planet had an Earth-like albedo (i.e. $\sim0.3$), there would have been no danger of desiccation while mode 1 was active. Only if initially $E_1$ were more than twice $E_2$, or if the albedo were less than half of Earth's, would we expect the inner planet to have been desiccated. There is no particular reason to expect mode 1 to have had an initially much smaller or much larger amplitude than mode 2, so it is unlikely that this example system would have had an initial $E_1$ large enough to dessicate the inner planet. 

On the other hand, if the system were in the portion of the long-term safe zone (the gray zone of Figure \ref{h1vsE2a2}), where $e_2 \simeq E_2 > 0.2$, the initial value of $E_1$ would have to be much smaller than the value of $E_2$ to avoid dessication. While such a situation cannot be ruled out, it seems likely that $E_1$ would have been large enough initially to result in dessication. 
Thus if planetary systems of the type under consideration here form with approximately equal initial eigenmode amplitudes, they are unlikely to have a habitable inner planet if $e_2 \gtrsim 0.2$. Any system within the remainder of the area of parameter space shown in gray in Figure \ref{h1vsE2a2} (with $e_2 \lesssim 0.2$) remains a candidate for habitability, even for the great ages of planetary systems orbiting small stars. If the inner planet initially had more water than assumed here, desiccation would have taken longer, so the acceptable range of parameters for habitability in Figure \ref{h1vsE2a2} would extend to even higher initial values of $e_2$.

\section{Discussion and Conclusions}\label{concl}

The presence of an outer companion to an Earth-mass planet orbiting in the classically defined habitable zone (HZ) of a low-mass star may induce enough tidal heating to maintain geophysical activity, and hence the possibility of life, even after other internal heat sources have died away. Figure \ref{h1vsE2a2} identifies (in gray shading) the portion of parameter space for the outer planet (mass, semi-major axis, eccentricity) for which the inner planet's tidal heating would be adequate, without over-heating and sterilizing it. Section \ref{evollt} shows that after the initial short-lived mode of the secular interaction dies away, this tidal heating likely remains steady for tens of billions of years. However, Section \ref{evolst} demonstrates that for cases with $e_2\gtrsim0.2$, the short-lived mode would probably have been so strong that the inner planet would have been desiccated very early in its history. Thus the range of acceptable parameters of the outer planet for habitability of the inner planet is given by the gray area in Figure \ref{h1vsE2a2}, without the portions with $e_2\gtrsim0.2$.

It is important to recognize that these boundaries of this area are very fuzzy (and are accordingly shown that way in Figure \ref{h1vsE2a2}). They may shift as we learn more about the criteria for supporting life and preventing sterilization. The acceptable range of parameters may accordingly shift somewhat to a different portion of Figure \ref{h1vsE2a2}. Among the many uncertainties that affect the boundaries are the geophysical criteria for initiation and maintenance of transport between the interior and surface, the degree to which life requires such transport, and the effective value of the tidal dissipation parameter $Q'$. The initial amount of water, if in excess of that of the Earth, could shift the upper limit on $e_2$ to a larger value as well.

The upper limit on the value of $e_2$ also depends on the assumption that the initial amplitudes $E_m$ of both secular eigenmodes are similar. If the orbital initial conditions are such that $E_2>E_1$, then the boundary would be more restrictive, moving to lower values of the current $e_2$. On the other hand, if initially $E_2<E_1$, then the early short-lived heat would have been less likely to desiccate the inner planet even if the current $e_2$ exceeds $\sim0.2$. In fact, the evaluation of observed multi-planet systems generally indicates considerable disparity in mode amplitudes within a given system \citep{VL12}. On the other hand, the variation in mode amplitudes within single systems could be artifacts of poorly defined observation-based current eccentricities and pericenter longitudes.

There are various qualitative issues that may also change the boundaries of the acceptable range of $m_2$, $a_2$, and $e_2$ values. We have ruled out systems with too large a value of $e_2$ because of the early desiccation, but if that heating phase ended early enough, it is conceivable that water was redelivered by subsequent accretion of icy material. Planet formation as a general process remains too poorly understood to exclude that possibility. 

Here we have considered very-low-mass stars ($\sim0.1$ solar mass).  For larger stars there would similarly be a range of parameters for an outer planet that would yield an adequate rate of tidal heating for long-term habitability. In such cases, the habitable zone is farther out so the inner (HZ) planet needs a larger eccentricity (Equation \ref{heat}). To meet that requirement, an outer planet of a given mass would also need a larger eccentricity than for the cases calculated here. For example, for a star of mass 0.3 times the Sun's, an outer Neptune-mass planet would need an eccentricity of about 0.2 [CHECK]. Note however that the theoretical framework we are using, second-order classical secular theory, breaks down for very high eccentricities.  Thus, while this mechanism for providing long-term tidal heating likely works at higher stellar masses, our analysis does not apply for stellar masses greater then $\sim0.3$ solar masses.

Our results above apply only to the specific inner-planet mass (Earth's), stellar mass (1/10 solar), and distance from the star chosen for our example. However, similar calculations for any other hypothetical or observed system are straightforward and can be carried out in a similar way to those demonstrated here.  Given the inner planet and stellar masses and separation, one can calculate its heating rate as a function of an outer planet's mass, semimajor axis and eccentricity $e_2$ as follows: Assuming that the fast-damping mode is gone, the ratio (equivalent to the eigenvector) of the inner planet's eccentricity, $e_1$, to $e_2$ is given by Equation 7b of \citet{Gre11} with appropriate care to the different notation. With $e_1$ determined in that way, the tidal heating of the inner planet can be computed from Equation \ref{heat} above. From such calculations, a contour plot equivalent to Figure \ref{h1vsE2a2} can be constructed.

\begin{figure*}
\includegraphics[width=\textwidth]{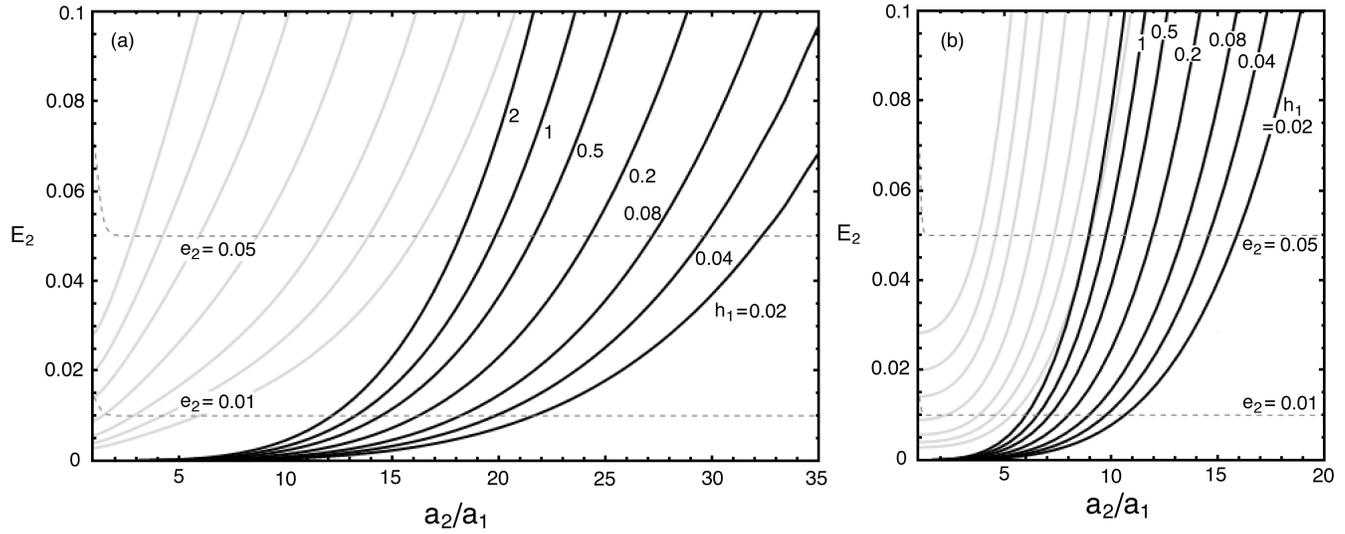}
\caption{Lines of constant $h_1$ versus the ratio of the planets' semi-major axes $a_2/a_1$ and the amplitude of mode 2 $E_2$. Similar to Figure \ref{h1vsE2a2} except that the abscissa is $a_2/a_1$ rather than $a_2$. The black lines are lines of constant $h_1$ for planets orbiting the star DEN1048. Grey lines are for planets orbiting the hypothetical late M dwarf discussed in Section \ref{ltinst}, Figure \ref{h1vsE2a2}. a) When $m_2 = 1$ Neptune mass, and b) when $m_2 = 1$ Earth mass.}
\label{DEN}
\end{figure*}

For example, consider the M dwarf DEN1048 with a mass of 0.07 solar masses and luminosity of $3.5\times10^{-6}$ \citep{Del01}. If an Earth-mass planet were discovered at the distance corresponding to the same instellation as received from the Earth ($2\times10^{-3}$ AU) , the heating rate as a function of the outer planet's eccentricity and semi-major axis would be as shown in Figure \ref{DEN}. As in Figure \ref{h1vsE2a2}, for any given assumption about the appropriate amount of heat to support life, this plot shows the orbital parameters of an outer planet that would deliver that amount of heat.

Although we have considered only two-planet systems here, systems with more planets could similarly alter habitability.  The calculations of secular interactions would be similar, although complicated by the greater number of eigenmodes.  More than one eigenmode could be long-lived, which would result in periodically varying eccentricities and tidal heating rates even after the rapidly damping modes have died out. If the geophysical response is slow compared with the period of the secular variations, the average heating rate would apply.  Otherwise there could be periodic changes in the habitability of the surface environment.

The examples presented here demonstrate that a reasonable fraction of terrestrial-scale planets in the HZ of very old, low-mass stars may be able to sustain life, even though without a satisfactory companion they would have cooled off by now.  The requirements on the outer planet are not extremely stringent.  For example, one could well imagine a Neptune-size outer planet a few times farther out than the rocky planet with an orbital eccentricity $\sim0.01-0.02$. Not only would such an outer planet yield an appropriate amount of tidal heating to allow life, but the heating would be at a steady rate for at least tens of Gyr.

An intriguing aspect of this investigation is that a planet orbiting a very low-mass M dwarf might be habitable for the duration of its main sequence lifetime of up to 10 trillion years \citep{Lau97}. Thus a planet orbiting in the HZ of a $\sim0.1$ solar mass star with an outer companion that sustains modest tidal heating could be the longest-lived surface habitat in the universe. While these small stars do not brighten appreciatively over the age of the universe, they will over the course of $10^{12}$ years, and so the planet would need to remain in the continuous HZ (see, e.g. \citealt{Rus13}) despite the slow decay of its orbit by tidal evolution. The Earth will become uninhabitable within the next few billion years, and so this type of planet may be the ultimate home for humanity in the very (very) distant future. Similarly, if these planets are the ideal locations for long-term survival, they may already be inhabited by space-faring civilizations and hence may be favorable targets for SETI.

With most observational techniques, Earth-scale planets are more readily discoverable within the HZs of old, low-mass stars than in the HZs of more massive stars. The presence of an appropriate outer planet may be critical to the actual habitability, and our results show that a substantial range of masses and orbits for the outer planet may well meet the requirements. Thus, a search plan for HZ Earth-scale planets should include follow-up searches for companions that could enhance the likelihood of habitability.

\section*{Acknowledgments}

Christa Van Laerhoven would like to acknowledge the NASA Earth and Space Science Fellowship program. Rory Barnes would like to acknowledge NSF grant AST-110882, and the NASA Astrobiology Institute's Virtual Planetary Laboratory lead team.


\label{lastpage}

\end{document}